\documentclass[aps,prl,twocolumn,superscriptaddress]{revtex4-2}
\usepackage[dvipdfmx]{graphicx}
\usepackage{color}
\usepackage{ulem}

\begin{document}

\title{Semiconducting Electronic Structure of the Ferromagnetic Spinel $\mathbf{Hg}\mathbf{Cr}_2\mathbf{Se}_4$\\ Revealed by Soft-X-Ray Angle-Resolved Photoemission Spectroscopy}

\author{Hiroaki~Tanaka}
\affiliation{Institute for Solid State Physics, The University of Tokyo, Kashiwa, Chiba 277-8581, Japan}

\author{Andrei~V.~Telegin}
\affiliation{M.N. Mikheev Institute of Metal Physics, UB RAS, Ekaterinburg 620108, Russia}

\author{Yurii~P.~Sukhorukov}
\affiliation{M.N. Mikheev Institute of Metal Physics, UB RAS, Ekaterinburg 620108, Russia}

\author{Vladimir~A.~Golyashov}
\affiliation{Institute of Semiconductor Physics, SB RAS, Novosibirsk 630090, Russia}
\affiliation{Synchrotron Radiation Facility SKIF, Boreskov Institute of Catalysis, SB RAS, Kol'tsovo 630559, Russia}

\author{Oleg~E.~Tereshchenko}
\affiliation{Institute of Semiconductor Physics, SB RAS, Novosibirsk 630090, Russia}
\affiliation{Synchrotron Radiation Facility SKIF, Boreskov Institute of Catalysis, SB RAS, Kol'tsovo 630559, Russia}

\author{Alexander~N.~Lavrov}
\affiliation{Nikolaev Institute of Inorganic Chemistry, SB RAS, Novosibirsk 630090, Russia}

\author{Takuya~Matsuda}
\affiliation{Institute for Solid State Physics, The University of Tokyo, Kashiwa, Chiba 277-8581, Japan}

\author{Ryusuke~Matsunaga}
\affiliation{Institute for Solid State Physics, The University of Tokyo, Kashiwa, Chiba 277-8581, Japan}

\author{Ryosuke~Akashi}
\affiliation{Quantum Materials and Applications Research Center, National Institutes for Quantum Science and Technology, Meguro-ku, Tokyo 152-0033, Japan}

\author{Mikk~Lippmaa}
\affiliation{Institute for Solid State Physics, The University of Tokyo, Kashiwa, Chiba 277-8581, Japan}

\author{Yosuke~Arai}
\affiliation{Institute for Solid State Physics, The University of Tokyo, Kashiwa, Chiba 277-8581, Japan}

\author{Shinichiro~Ideta}
\altaffiliation[Present address: ]{Hiroshima Synchrotron Radiation Center, Hiroshima University, Higashi-hiroshima, Hiroshima 739-0046, Japan.}
\affiliation{UVSOR Facility, Institute for Molecular Science, Okazaki, Aichi 444-8585, Japan}

\author{Kiyohisa~Tanaka}
\affiliation{UVSOR Facility, Institute for Molecular Science, Okazaki, Aichi 444-8585, Japan}

\author{Takeshi~Kondo}
\affiliation{Institute for Solid State Physics, The University of Tokyo, Kashiwa, Chiba 277-8581, Japan}
\affiliation{Trans-scale Quantum Science Institute, The University of Tokyo, Bunkyo-ku, Tokyo 113-0033, Japan}

\author{Kenta~Kuroda}
\email{kuroken224@hiroshima-u.ac.jp}
\affiliation{Graduate School of Advanced Science and Engineering, Hiroshima University, Higashi-hiroshima, Hiroshima 739-8526, Japan}
\affiliation{International Institute for Sustainability with Knotted Chiral Meta Matter (WPI-SKCM${}^{2}$), Hiroshima University, Higashi-hiroshima, Hiroshima 739-8526, Japan}

\date{\today}

\begin{abstract}
We study the electronic structure of the ferromagnetic spinel $\mathrm{Hg}\mathrm{Cr}_2\mathrm{Se}_4$ by soft-x-ray angle-resolved photoemission spectroscopy (SX-ARPES) and first-principles calculations.
While a theoretical study has predicted that this material is a magnetic Weyl semimetal, SX-ARPES measurements give direct evidence for a semiconducting state in the ferromagnetic phase.
Band calculations based on the density functional theory with hybrid functionals reproduce the experimentally determined band gap value, and the calculated band dispersion matches well with ARPES experiments.
We conclude that the theoretical prediction of a Weyl semimetal state in $\mathrm{Hg}\mathrm{Cr}_2\mathrm{Se}_4$ underestimates the band gap, and this material is a ferromagnetic semiconductor.
\end{abstract}

\maketitle

Angle-resolved photoemission spectroscopy (ARPES)  is a powerful tool for investigating the electronic structure of solids because ARPES can directly observe band dispersions \cite{RevModPhys.93.025006}.
By comparing ARPES spectra with \textit{ab initio} band calculations, it is possible to verify the presence of various emergent electronic properties, such as Weyl points in polar Weyl semimetals \cite{doi:10.1126/science.aaa9297,Lv2015,Yang2015,PhysRevLett.122.176402} and magnetic Weyl semimetals \cite{Kuroda2017,Liu2019}.
Density functional theory (DFT) \cite{PhysRev.136.B864,PhysRev.140.A1133} calculations are often used to predict the electronic structure of a material, but DFT calculations are known to underestimate the width of the band gap of solids \cite{AULBUR20001}.
The discrepancy between calculations and experiments can be eliminated with the help of hybrid functionals, where model parameters are adjusted to reproduce the experimentally determined materials' characteristics \cite{Marsman_2008}.

Among the materials showing very different characteristics in calculations and experiments, the ferromagnetic spinel $\mathrm{Hg}\mathrm{Cr}_2\mathrm{Se}_4$ (space group $Fd\bar{3}m$, $T_C\simeq 110\ \mathrm{K}$) \cite{PhysRevLett.15.493,PhysRev.151.367,doi:10.1143/JPSJ.31.123,SELMI1986121,BEBENIN2015127} is one of the most intriguing.
Magnetotransport studies have reported that this material is an insulator \cite{PhysRevLett.15.493} or either a $p$- or $n$-type semiconductor \cite{doi:10.1063/1.324928},  depending on crystal growth conditions.
Infrared optical measurements have observed a band gap of 0.3--0.8 eV depending on temperature \cite{LEHMANN1969965,doi:10.1143/JPSJ.34.68,Auslender1989}.
The proposed $s$-$d$ model \cite{AUSLENDER1989761} accounting for direct alllowed transition at the $\Gamma$ point and the band gap could explain the temperature dependence of electron mass and the anisotropy of the magnetoresistance of $p$-type materials \cite{https://doi.org/10.1002/pssb.2221580131}.
On the other hand, recent DFT calculations using the generalized gradient approximation (GGA) for the exchange-correlation functional have predicted a semimetallic electronic structure and the presence of magnetic Weyl points near the Fermi level \cite{PhysRevLett.107.186806}.
Their prediction and another study on pyrochlore iridates \cite{PhysRevB.83.205101} are two of the earliest works of magnetic Weyl semimetals.
Following the magnetic Weyl semimetal proposal, additional support for the Weyl semimetal state was found by the observation of the anomalous Hall effect (AHE) \cite{PhysRevLett.115.087002}.
However, the AHE of $\mathrm{Hg}\mathrm{Cr}_2\mathrm{Se}_4$ has an unconventional temperature dependence \cite{PhysRevLett.123.096601}.
Since then, a theoretical study has suggested that the observed AHE temperature dependence may be extrinsic, caused by scattering, and unrelated to magnetic Weyl semimetals \cite{PhysRevLett.124.156802}.
In addition, another report on the band dispersion of $\mathrm{Hg}\mathrm{Cr}_2\mathrm{Se}_4$ \cite{Guo_2012} has claimed that the modified Becke-Johnson exchange-correlation functional \cite{PhysRevLett.102.226401} yields a gapped magnetic semiconductor phase.
To the best of our knowledge, the direct observation of the band dispersion by ARPES has been missing, and the real electronic structure of $\mathrm{Hg}\mathrm{Cr}_2\mathrm{Se}_4$ is still under debate.

In this Letter, we report the first observation of the band dispersion of $\mathrm{Hg}\mathrm{Cr}_2\mathrm{Se}_4$ by soft-x-ray (SX) ARPES.
We observed only hole bands around the Fermi level and did not find any electron bands crossing them.
This result unambiguously indicates that $\mathrm{Hg}\mathrm{Cr}_2\mathrm{Se}_4$ has a semiconducting electronic structure in the ferromagnetic phase, which is consistent with the previously reported transport and optical properties \cite{PhysRevLett.15.493,doi:10.1063/1.324928,LEHMANN1969965,doi:10.1143/JPSJ.34.68}.
Our accurate DFT calculations using hybrid functionals could reproduce the semiconducting band dispersion, and the calculated band dispersion agrees well with the experimental results.
Our results highlight that a simple comparison of ARPES spectra with DFT calculations has to be handled with care.

\begin{figure}
\includegraphics{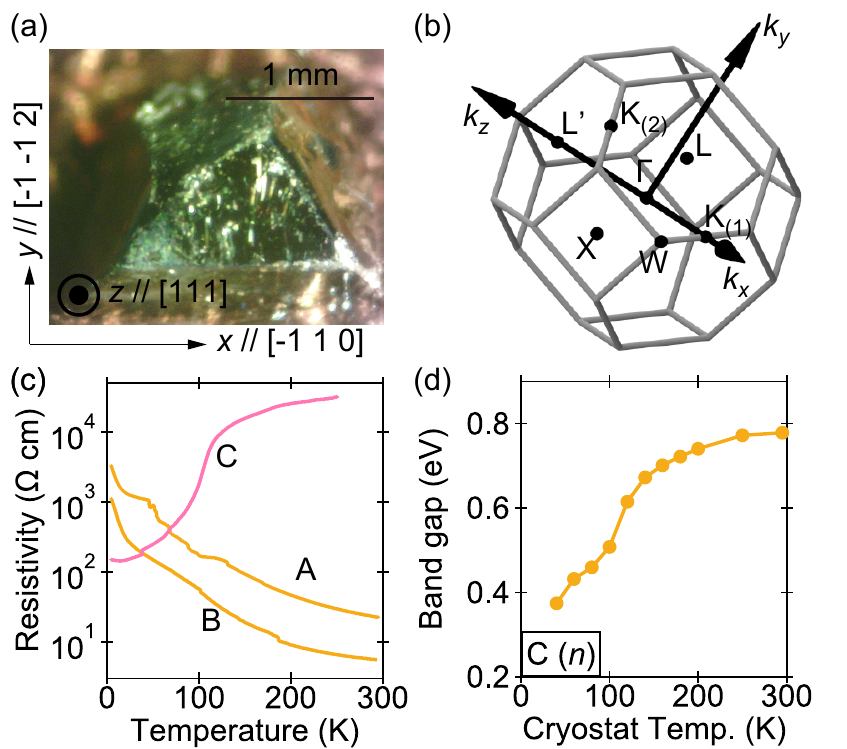}
\caption{\label{Fig: Basic_properties} Structure, transport properties, and optical properties. (a) Microscope image of an octahedral single crystal. (b) Brillouin zone and high-symmetric points. (c) Temperature dependence of the resistivity. (d) Temperature dependence of the band gap of crystal $C$ ($n$-type) estimated from the transmittance spectra.}
\end{figure}

Single crystals of $\mathrm{Hg}\mathrm{Cr}_2\mathrm{Se}_4$ were grown by the chemical transport reaction technique using $\mathrm{Cr}\mathrm{Cl}_3$ as a carrier agent.
The crystal structure was analyzed by powder x-ray diffraction (Supplemental Material Fig.\ S1 in \cite{Supplemental}).
\nocite{VANSETTEN201839}
\nocite{PhysRevB.97.121110}
We used three single crystals in the experiments; two of them ($A$ and $B$) were $p$-type, and the other ($C$) was $n$-type.
Since as-grown $\mathrm{Hg}\mathrm{Cr}_2\mathrm{Se}_4$ crystals usually become $p$-type, the $n$-type crystals were obtained by annealing them in a Hg vapor \cite{doi:10.1063/1.324928, SELMI1986121,SELMI19801285, Solin2000, Solin2008}.
A triangular face of a single crystal and the edge of the triangle [Fig.\ \ref{Fig: Basic_properties}(a)] correspond to the $(111)$ plane and the $[\bar{1}10]$ direction, respectively (Supplemental Material Note 2 in \cite{Supplemental}).
We set the $xyz$ axes so that the $x$ axis was parallel to the edge and the $z$ axis was perpendicular to the plane [Figs.\ \ref{Fig: Basic_properties}(a) and (b)].
The characteristics of $\mathrm{Hg}\mathrm{Cr}_2\mathrm{Se}_4$ single crystals were examined by transport, magnetic, and infrared transmission analysis (Note 3 in the Supplemental Material \cite{Supplemental}).
SX- and VUV-ARPES measurements were performed at BL25SU of SPring-8 \cite{Muro:ok5049}, and supplemental ARPES measurements using vacuum ultraviolet (VUV) light were performed at BL5U of UVSOR, respectively.
We used 400--550 eV SX light in SX-ARPES measurements and 60 eV VUV light in VUV-ARPES measurements.
The measurement temperature was kept around 50 K.
The samples were cleaved \textit{in situ} in ultrahigh vacuum better than $\sim 3\times10^{-8}\ \mathrm{Pa}$.
The cleaved surfaces were later examined by contact-mode atomic force microscopy (AFM) in air.
The electronic band structures were calculated from first principles using the Quantum Espresso \cite{Giannozzi_2017} and Wannier90 \cite{Pizzi_2020} packages (Note 3 in the Supplemental Material \cite{Supplemental}).
The charge density was calculated with the Kohn-Sham equation with the Heyd-Scuseria-Ernzerhof (HSE) 06 exchange-correlation functional \cite{doi:10.1063/1.1564060, doi:10.1063/1.2204597}.
After that, the Kohn-Sham bands were derived with the Wannier interpolation method \cite{PhysRevB.56.12847, PhysRevB.65.035109}.
The experimentally observed crystal structure was adopted, and spin-orbit coupling was included in the calculations.

We distinguished $p$- and $n$-types by the temperature dependence of the resistivity [Fig.\ \ref{Fig: Basic_properties}(c)]; $p$-type crystals ($A$ and $B$) exhibited increasing resistivity at lower temperature while the resistivity of an $n$-type crystal ($C$) dropped upon cooling at around $T_C\simeq 110\ \mathrm{K}$ \cite{doi:10.1063/1.324928,SELMI19801285,PhysRevLett.115.087002,PhysRevLett.123.047201}.
These conductivity types are consistent with the Seebeck coefficient signs determined by a hot-probe method at room temperature.
We observed some steps and spikes in the $\rho$--$T$ curves.
We think they can be due to the magnetic domain structure reformation with changing the temperature; the anomalous magnetoresistance behavior [Fig.\ S3(a) in the Supplemental Material \cite{Supplemental}] indicates unconventional transport properties at the magnetic domain boundaries.
Since the resistivity curve for the $n$-type sample is similar to that of metal, $n$-type $\mathrm{Hg}\mathrm{Cr}_2\mathrm{Se}_4$ has been considered to be a degenerated semiconductor \cite{doi:10.1063/1.324928,SELMI19801285}.
The magnetic property measurements showed ferromagnetic behavior below $T_\mathrm{C}$ [Figs.\ S3(b) and S3(c) in the Supplemental Material \cite{Supplemental}].
The transmittance spectra of $n$-type $\mathrm{Hg}\mathrm{Cr}_2\mathrm{Se}_4$ (crystal $C$) showed a steplike feature [Fig.\ S4 in the Supplemental Material \cite{Supplemental}]; the transmittance steeply dropped to nearly zero above a threshold, indicating the existence of a band gap.
The band gap energy estimated from these spectra [Fig.\ \ref{Fig: Basic_properties}(d)] exhibits a similar temperature dependence to previous studies \cite{LEHMANN1969965,doi:10.1143/JPSJ.34.68,Auslender1989}.
Since the previous studies use $p$-type samples to measure the band gap, our measurements of the $n$-type sample clarify that the band gap behavior is independent of the carrier type.
All these results are consistent with the previous reports claiming that $\mathrm{Hg}\mathrm{Cr}_2\mathrm{Se}_4$ is a ferromagnetic semiconductor.

\begin{figure}
\includegraphics{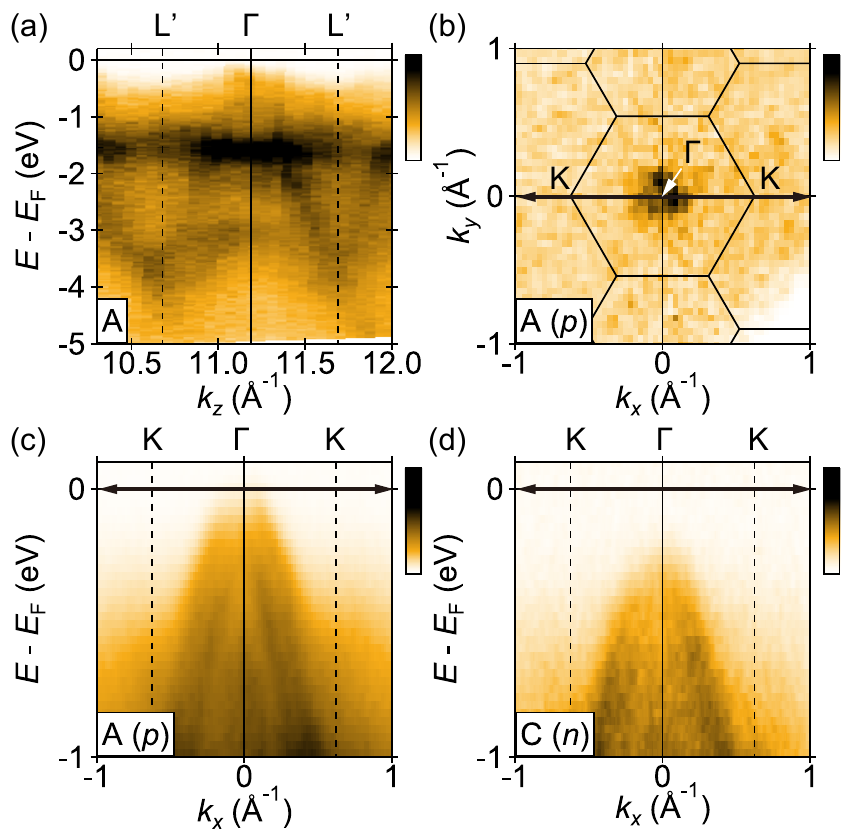}
\caption{\label{Fig: SX-ARPES} SX-ARPES measurements. (a) $k_z$ dispersion along the $\Gamma L^\prime$ direction. (b) Fermi surface of crystal $A$ ($p$-type). Black solid lines represent the cross section of the Brillouin zone. (c), (d) Band dispersions of crystals $A$ ($p$-type) and $C$ ($n$-type) along the $\Gamma K$ direction [black arrow in (b)]. (b)-(d) were taken with 475 eV SX light.}
\end{figure}

Using SX-ARPES, we present conclusive evidence for the semiconducting electronic structure of $\mathrm{Hg}\mathrm{Cr}_2\mathrm{Se}_4$ in the ferromagnetic phase.
We could observe band dispersions from all three samples.
Figure \ref{Fig: SX-ARPES}(a) represents the $k_z$ dispersion of $\mathrm{Hg}\mathrm{Cr}_2\mathrm{Se}_4$ (Crystal A) along the $\Gamma L^\prime$ direction obtained by changing the photon energy from 400 eV to 550 eV.
We used the inner potential of 12 eV and the work function of 5 eV to convert the photon energy to the wavevector.
These values gave properly symmetric $k_z$ dispersions about the $\Gamma$ or $L^\prime$ points.

The $k_z$ dispersion analysis revealed that the 475 eV SX light captures the bulk band dispersion around the $\Gamma$ point.
Figure \ref{Fig: SX-ARPES}(b) shows the Fermi surface of $p$-type $\mathrm{Hg}\mathrm{Cr}_2\mathrm{Se}_4$, and Figs.\ \ref{Fig: SX-ARPES}(c) and (d) show the $\Gamma K$ direction band dispersions of $p$- and $n$-type $\mathrm{Hg}\mathrm{Cr}_2\mathrm{Se}_4$, respectively.
All dispersions exhibit only hole bands around the $\Gamma$ point.
The top of the hole bands in the $p$-type sample [Fig.\ \ref{Fig: SX-ARPES}(c)] is close to the Fermi level, which is consistent with the $p$-type carrier of the sample.
On the other hand, the hole bands in the $n$-type sample [Fig.\ \ref{Fig: SX-ARPES}(d)] are at a slightly lower energy than those of the $p$-type sample [Fig.\ \ref{Fig: SX-ARPES}(c)], reflecting the electron doping of the $n$-type sample.
Note 6 in the Supplemental Material \cite{Supplemental} discusses the energy level of the $n$-type sample obtained by the infrared absorption and ARPES measurements.
While earlier band calculation \cite{PhysRevLett.107.186806} has predicted that the hole and electron bands would cross around the $\Gamma$ point in the ferromagnetic phase, such electron bands were absent in the experimental results.
Thus we conclude that $\mathrm{Hg}\mathrm{Cr}_2\mathrm{Se}_4$ is a semicondutor in the ferromagnetic phase, not a magnetic Weyl semimetal in constrast to the theoretical prediction \cite{PhysRevLett.107.186806}.

Although SX-ARPES measurements were successful, ARPES measurements using VUV light, which is typically used in studies of Weyl semimetals, could not detect any dispersion around the Fermi level (Note 6 in the Supplemental Material \cite{Supplemental}).
The difference in SX- and VUV-ARPES results is caused by the rough cleaved surface of the spinel $\mathrm{Hg}\mathrm{Cr}_2\mathrm{Se}_4$.
We will argue this point later.

\begin{figure} 
\includegraphics{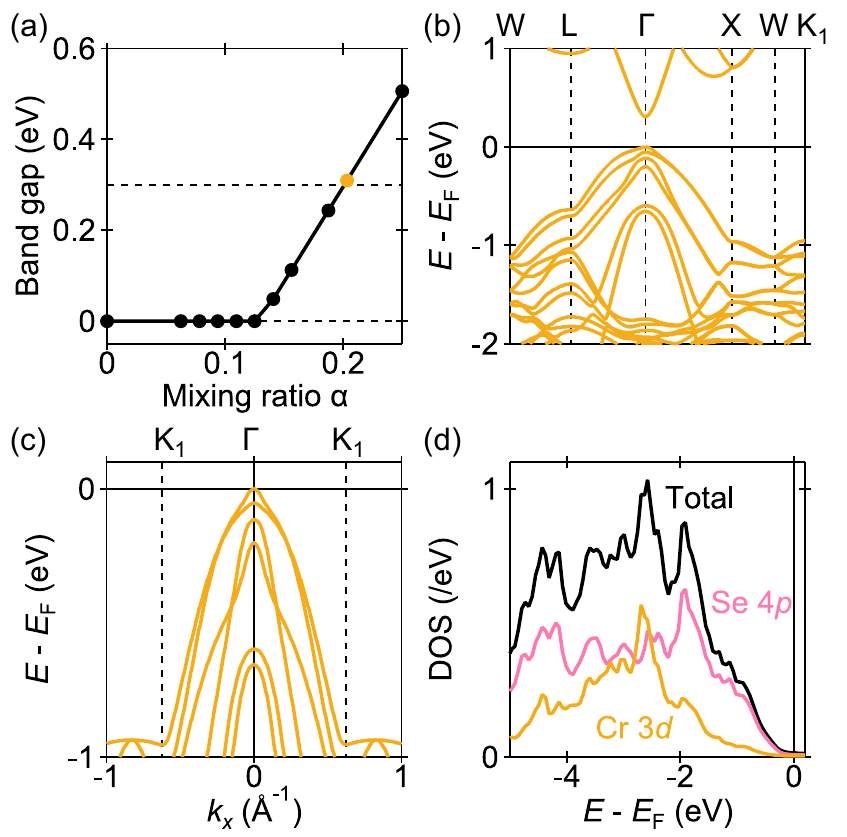}
\caption{\label{Fig: DFT} DFT calculations. (a) Relation between the mixing ratio of Hartree-Fock exchange and the band gap. The horizontal dashed line at 0.3 eV represents the experimental band gap at low temperatures \cite{doi:10.1143/JPSJ.34.68}. (b), (c) Band dispersions at the mixing ratio corresponding to the orange point in (a). (d) Density of states (DOS) of the valence bands.}
\end{figure}

We further analyzed the electronic structure of $\mathrm{Hg}\mathrm{Cr}_2\mathrm{Se}_4$ by DFT calculations with the HSE06 hybrid functional.
In the HSE06 functional, $\alpha$ represents the mixing ratio of the Hartree-Fock exact exchange into the Perdew-Burke-Ernzerhof (PBE) exchange-correlation functional \cite{doi:10.1063/1.472933} of the GGA type; $\alpha=0$ corresponds to the original PBE functional, and $\alpha=0.25\equiv\alpha_{\mathrm{HSE06}}$ is the standard HSE06 value.
The PBE functional gives a semimetallic band structure with magnetic Weyl points near the Fermi level \cite{PhysRevLett.107.186806}.
However, the standard local density approximation and GGA exchange-correlation functionals often underestimate the band gaps of semiconductors, even occasionally yielding incorrect gap closing.
That is why the exact exchange is included in hybrid functionals to mitigate this problem \cite{Marsman_2008}.
Recent first-principles examinations \cite{https://doi.org/10.1002/pssb.201046195,PhysRevB.89.195112} have suggested that the optimum value of $\alpha$ for accurate band gap prediction may be smaller in narrow-gap semiconductors. To see the robustness of the gap opening, we, therefore, calculated the electronic structure with various $\alpha$ between 0 and 0.25.
Figure \ref{Fig: DFT}(a) shows the band gaps for different $\alpha$ values.
A nonzero mixing ratio $\alpha$ yields a gapped electronic structure for $\alpha> 0.12$ and the band gap increases with increasing $\alpha$ [Fig.\ \ref{Fig: DFT}(a)].

Here we show the calculated band structure with $\alpha=\alpha_{\mathrm{HSE06}}\times13/16$ [orange point in Fig.\ \ref{Fig: DFT}(a)], where the band gap (0.31 eV) was approximately the same as the experimental one at low temperatures  \cite{doi:10.1143/JPSJ.34.68}; as discussed above, the band gap size depends on the temperature but is independent of the carrier type. Note 7 in the Supplemental Material \cite{Supplemental} represents the calculation result with $\alpha=\alpha_\mathrm{HSE06}$.
Figure \ref{Fig: DFT}(c) shows the band dispersion along the $k_x$ direction, which agrees well with the experimental results [Figs.\ \ref{Fig: SX-ARPES}(c) and (d)].
While the magnetic moments of the calculated electronic structure are aligned along the vertical direction in Fig.\ \ref{Fig: Basic_properties}(b), the symmetry lowering due to the ferromagnetism is negligible (Note 8 in the Supplemental Material \cite{Supplemental}).
The calculated density of states [Fig.\ \ref{Fig: DFT}(d)] shows the contribution of Cr $3d$ and Se $4p$ orbitals to the valence band, and the contribution of Cr $3d$ orbitals is consistent with our resonant PES measurements (Note 9 in the Supplemental Material \cite{Supplemental}).
Our DFT calculations exhibit that the hybrid functional with the appropriate mixing ratio reproduces both the band gap determined by infrared spectra and the band dispersion obtained by SX-ARPES.

As was noted earlier, the roughness of the cleaved surface of the spinel $\mathrm{Hg}\mathrm{Cr}_2\mathrm{Se}_4$ affects photoemission spectra particularly in VUV-ARPES.
The followinig argument about the rough surface illustrates the benefit of using SX-ARPES to observe the band dispersion of a material that does not have a good cleaving plane.

\begin{figure}
\includegraphics{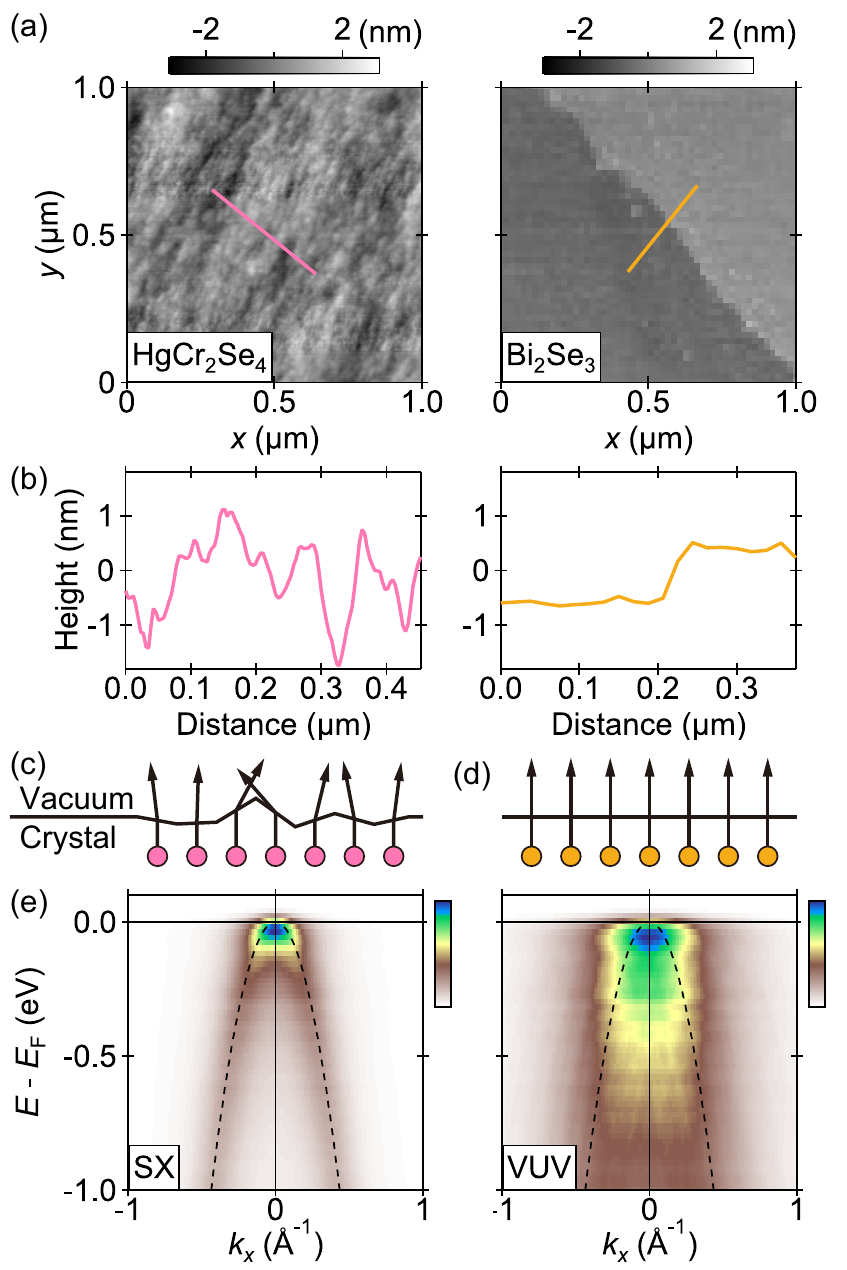}
\caption{\label{Fig: Surface} Effect of surface roughness on ARPES spectra. (a) AFM images of $\mathrm{Hg}\mathrm{Cr}_2\mathrm{Se}_4$ and $\mathrm{Bi}_2\mathrm{Se}_3$. Both panels use the same color scale. (b) Height profiles along the pink and orange lines in (a). (c), (d) Schematics of a rough surface modeled by randomly-oriented flat surfaces and a flat surface.  (e) Simulated photoemission spectra refracted by a rough surface in SX and VUV regions. The dashed parabolas represent the band dispersion before the refraction.}
\end{figure}

The AFM images in Figs.\ \ref{Fig: Surface}(a) and \ref{Fig: Surface}(b) compare the cleaved surfaces of $\mathrm{Hg}\mathrm{Cr}_2\mathrm{Se}_4$ and $\mathrm{Bi}_2\mathrm{Se}_3$, which is a quintuple-layered material hosting topological surface states \cite{Zhang2009,Xia2009}.
The cleaved surface of $\mathrm{Hg}\mathrm{Cr}_2\mathrm{Se}_4$ is indeed rough, while $\mathrm{Bi}_2\mathrm{Se}_3$ exhibits large flat terraces with occasional steps, reflecting the layered structure.
Compared to the flat surface assumed in typical ARPES measurements, the uneven surface of $\mathrm{Hg}\mathrm{Cr}_2\mathrm{Se}_4$ will create an unusual situation when photoelectrons escape from the crystal to the vacuum.
Although the electronic structure of the rough surface region can be greatly different from that of the bulk, the photoemission from such an incoherent electronic structure is expected to be very weak because, roughly speaking, the photoemission probability is determined by the matrix element of the ground state and the plane wave.
Therefore we think our ARPES spectra come from the photoemission in the deeper region [Fig.\ S10 in the Supplemental Material \cite{Supplemental}], in which the electronic structure is sufficiently related to the bulk.
The clear $k_z$ dispersion [Fig.\ \ref{Fig: SX-ARPES}(a)] and the coincidence with the calculated bulk band dispersion [Fig.\ \ref{Fig: DFT}(c)] support our argument.

In the three-step model \cite{RevModPhys.93.025006}, when a photoelectron escape from the crystal, the out-of-plane component of the photoelectron wavevector is modified while the in-plane component is unchanged [Fig.\ S11 in the Supplemental Material \cite{Supplemental}].
While standard ARPES studies discuss the refraction of the photoelectron using a flat plane as shown in Fig.\ \ref{Fig: Surface}(d), such an analysis cannot be applied straightforwardly to $\mathrm{Hg}\mathrm{Cr}_2\mathrm{Se}_4$ with a rough cleaved surface.
We simulated the photoemission from a rough surface as a sum of photoemission spectra from randomly-oriented flat surfaces [Fig.\ \ref{Fig: Surface}(c) and Ref.\ \cite{Refraction}] using the hole band dispersion in Fig.\ \ref{Fig: SX-ARPES}(c).
The simulated spectra in the SX and VUV regions are shown in Fig.\ \ref{Fig: Surface}(e) for photon energies that are approximately the same as in the experiments.
A slightly broadened parabolic dispersion is retained in the SX region.
However, the simulated spectra in the VUV region are too broad to see a dispersion curve.
The constant energy maps also exhibit the similar behavior [Fig.\ S12 in the Supplemental Material \cite{Supplemental}].
This difference arises because the wavevector of a photoelectron in the VUV region is shorter than that in the SX region and, therefore, more strongly refracted [Fig.\ S13 in the Supplemental Material \cite{Supplemental}].
A clear advantage of SX-ARPES over VUV-ARPES is thus that SX-ARPES can be used to study materials that are not easy to cleave and therefore suffer from increased surface roughness.
We note that the bulk sensitivity of SX-ARPES \cite{STROCOV20191} and a micro-focused beam of a dedicated SX-ARPES beamline \cite{Muro:ok5049} can also contribute to the clear dispersion observed in the SX-ARPES experiments.

In conclusion, we show for the first time the ARPES result of the ferromagnetic spinel $\mathrm{Hg}\mathrm{Cr}_2\mathrm{Se}_4$, which is direct evidence for the semiconducting electronic structure of this material.
SX-ARPES was used to visualize the valence band dispersion of $\mathrm{Hg}\mathrm{Cr}_2\mathrm{Se}_4$.
We only observed hole bands while electron bands were absent below the Fermi level.
These ARPES results are consistent with the existence of a band gap determined by infrared transmittance spectra.
DFT calculations with the HSE06 hybrid functional reproduced the gapped band dispersion.
We adjusted the mixing ratio of the HSE06 functional so that the calculations gave the experimentally observed band gap, and obtained dispersion curves that agreed well with our ARPES result.
Our study draws attention to choosing appropriate functionals in DFT calculations and the advantage of SX-ARPES in investigating the electronic structure of solid crystals.

\begin{acknowledgments}
We thank Y.\ Ishida for supporting the analysis of ARPES data \cite{Ishida2018}, and P. Zhang, S. Sakuragi, and R. Noguchi for supporting our SX-ARPES experiments.
A.V.T. thanks the Brain Pool (BP) Program funded by the Ministry of Science and ICT (MSIT) (No.\ 2021H1D3A2A01096552).
V.A.G. and O.E.T. acknowledge support from the Russian Science Foundation (Grant No.\ 22-12-20024) and the government of the Novosibirsk region (p-9).
K.K. acknowledges support from the Murata Science Foundation and the Futaba research grant program.
This work is also supported by Grant-in-Aid for JSPS Fellows (Grant No.\ JP21J20657), Grant-in-Aid for Early-Career Scientists (Grant No.\ JP21K13858), Grant-in-Aid for Scientific Research on Innovative Areas (Grant No. JP22H04483), Grant-in-Aid for Scientific Research (B) (Grant No.\ JP22H01943) Grant-in-Aid for Scientific Research (A) (Grants No.\ JP21H04652 and No.\ JP21H04439), Japan Science and Technology Agency (JST) (Grant No.\ JPMJMI21G2), JST, PRESTO (Grant No.\ JPMJPR20LA), program of Ministry of Education and Science of the Russian Federation (``Spin'' No.\ 122021000036-3), Quantum Leap Flagship Program from Ministry of Education, Culture, Sports, Science and Technology (MEXT Q-LEAP) (Grant No.\ JPMXS0118068681), and Photon and Quantum Basic Research Coordinated Development Program from MEXT.
Transmission spectroscopy and Laue back-reflection measurements  were performed using the facilities of the Materials Design and Characterization Laboratory in the Institute for Solid State Physics (ISSP), the University of Tokyo.
Some of the calculations were performed at the Supercomputer Center in ISSP, the University of Tokyo.
The synchrotron radiation experiments were performed with the approval of UVSOR (Proposals No.\ 20-757 and No.\ 20-842) and JASRI (Proposals No.\ 2019B1092 and No.\ 2020A0606).
\end{acknowledgments}
\bibliography{HgCr2Se4_SX-ARPES_references}

\end{document}


\title{Supplemental Material: Semiconducting Electronic Structure of the Ferromagnetic Spinel $\mathbf{Hg}\mathbf{Cr}_2\mathbf{Se}_4$ Revealed by Soft-X-Ray Angle-Resolved Photoemission Spectroscopy}

\author{Hiroaki~Tanaka}
\affiliation{Institute for Solid State Physics, The University of Tokyo, Kashiwa, Chiba 277-8581, Japan}

\author{Andrei~V.~Telegin}
\affiliation{M.N. Mikheev Institute of Metal Physics, UB RAS, Ekaterinburg 620108, Russia}

\author{Yurii~P.~Sukhorukov}
\affiliation{M.N. Mikheev Institute of Metal Physics, UB RAS, Ekaterinburg 620108, Russia}

\author{Vladimir~A.~Golyashov}
\affiliation{Institute of Semiconductor Physics, SB RAS, Novosibirsk 630090, Russia}
\affiliation{Synchrotron Radiation Facility SKIF, Boreskov Institute of Catalysis, SB RAS, Kol'tsovo 630559, Russia}

\author{Oleg~E.~Tereshchenko}
\affiliation{Institute of Semiconductor Physics, SB RAS, Novosibirsk 630090, Russia}
\affiliation{Synchrotron Radiation Facility SKIF, Boreskov Institute of Catalysis, SB RAS, Kol'tsovo 630559, Russia}

\author{Alexander~N.~Lavrov}
\affiliation{Nikolaev Institute of Inorganic Chemistry, SB RAS, Novosibirsk 630090, Russia}

\author{Takuya~Matsuda}
\affiliation{Institute for Solid State Physics, The University of Tokyo, Kashiwa, Chiba 277-8581, Japan}

\author{Ryusuke~Matsunaga}
\affiliation{Institute for Solid State Physics, The University of Tokyo, Kashiwa, Chiba 277-8581, Japan}

\author{Ryosuke~Akashi}
\affiliation{Quantum Materials and Applications Research Center, National Institutes for Quantum Science and Technology, Meguro-ku, Tokyo 152-0033, Japan}

\author{Mikk~Lippmaa}
\affiliation{Institute for Solid State Physics, The University of Tokyo, Kashiwa, Chiba 277-8581, Japan}

\author{Yosuke~Arai}
\affiliation{Institute for Solid State Physics, The University of Tokyo, Kashiwa, Chiba 277-8581, Japan}

\author{Shinichiro~Ideta}
\affiliation{UVSOR Facility, Institute for Molecular Science, Okazaki, Aichi 444-8585, Japan}

\author{Kiyohisa~Tanaka}
\affiliation{UVSOR Facility, Institute for Molecular Science, Okazaki, Aichi 444-8585, Japan}

\author{Takeshi~Kondo}
\affiliation{Institute for Solid State Physics, The University of Tokyo, Kashiwa, Chiba 277-8581, Japan}
\affiliation{Trans-scale Quantum Science Institute, The University of Tokyo, Bunkyo-ku, Tokyo 113-0033, Japan}

\author{Kenta~Kuroda}
\affiliation{Graduate School of Advanced Science and Engineering, Hiroshima University, Higashi-hiroshima, Hiroshima 739-8526, Japan}
\affiliation{International Institute for Sustainability with Knotted Chiral Meta Matter (WPI-SKCM${}^{2}$), Hiroshima University, Higashi-hiroshima, Hiroshima 739-8526, Japan}

\date{\today}

\maketitle

\renewcommand{\thefigure}{S\arabic{figure}}
\renewcommand{\thetable}{S\Roman{table}}
\titleformat*{\section}{\bfseries}

\section{Note 1: Powder x-ray diffraction analysis}
\begin{figure}[hbt]
\includegraphics[width=160mm]{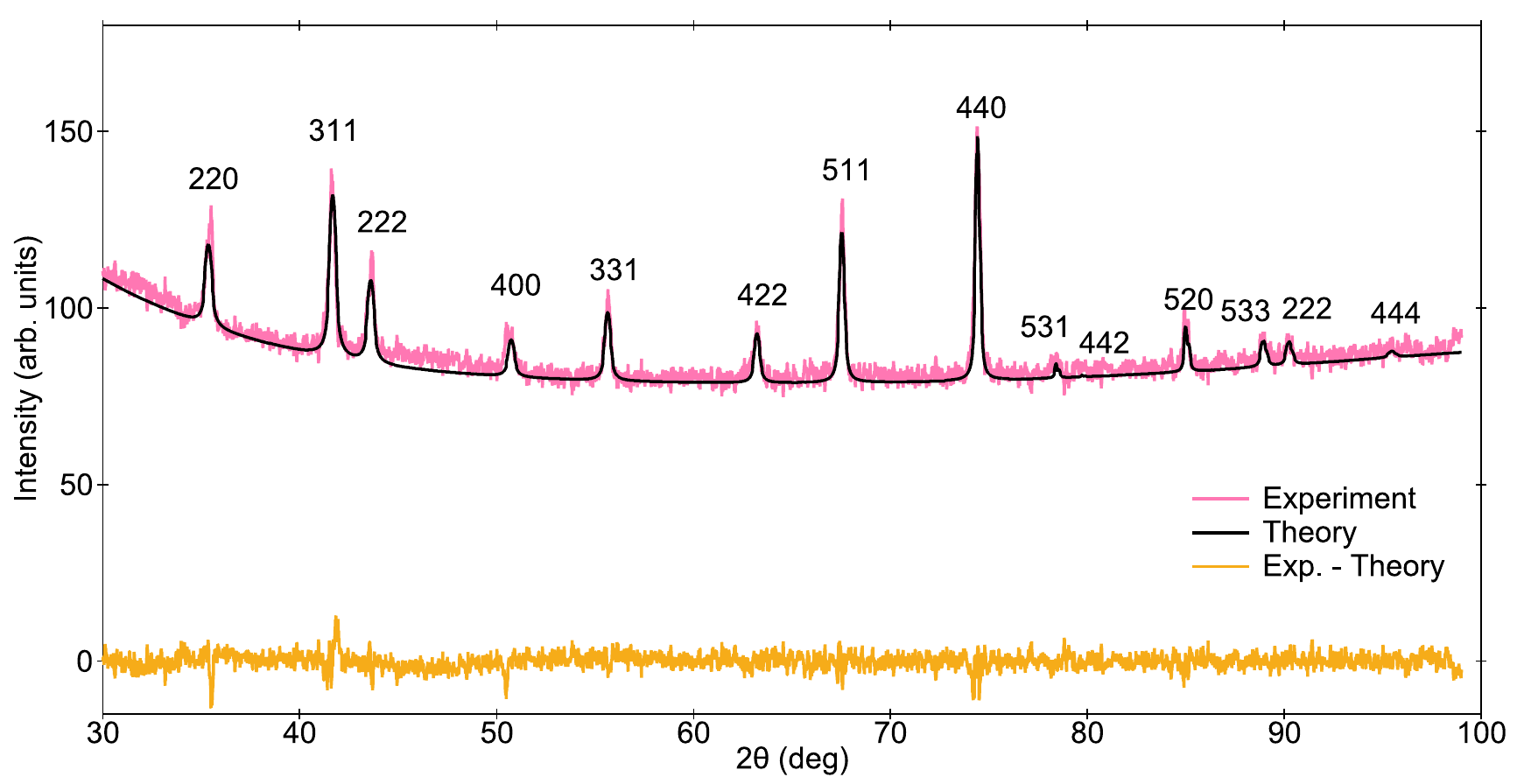}
\caption{\label{Fig: Powder_XRD} Powder x-ray diffraction spectra of $\mathrm{Hg}\mathrm{Cr}_2\mathrm{Se}_4$.}
\end{figure}

The crystal structure was analyzed  by powder x-ray diffraction using the Cr K$\alpha$ x-ray source.
Figure \ref{Fig: Powder_XRD} shows the analysis result.
The measurement result corresponds well with theoretical calculations.
The lattice constant was determined to be $a=10.749(8)\ \text{\AA}$, consistent with previous studies.

\clearpage

\section{Note 2: Laue back-reflection analysis}
\begin{figure}[hbt]
\includegraphics[width=160mm]{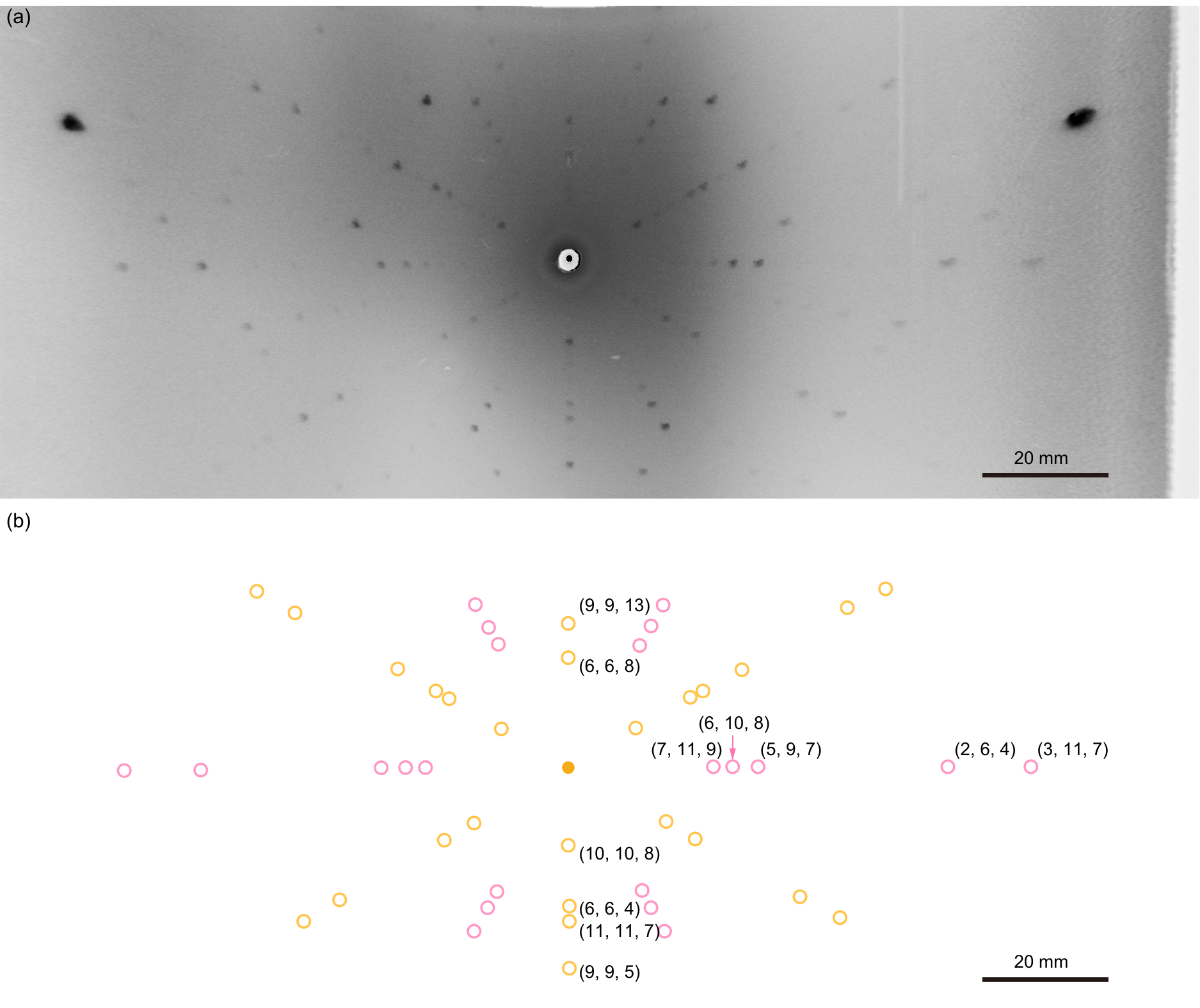}
\caption{\label{Fig: Laue} Laue back-reflection analysis. (a) Laue image of a $\mathrm{Hg}\mathrm{Cr}_2\mathrm{Se}_4$ single crystal. The orientation is the same as in Fig.\ 1(a). (b) Laue indices for intense spots.}
\end{figure}

We performed Laue back-reflection measurements using a Laue camera system (IPX-LC, IPX Co., Ltd.) with a tungsten anode x-ray tube to determine the crystal orientation; the tube potential was set to 30 keV.
Figure \ref{Fig: Laue}(a) shows the Laue image of a $\mathrm{Hg}\mathrm{Cr}_2\mathrm{Se}_4$ single crystal; we set a triangular plane parallel to the screen and the distance between the screen and the crystal was 60 mm.
The pattern exhibits three-fold rotational symmetry, indicating that the plane is perpendicular to the $[111]$ axis.
We calculated the Laue indices for intense spots as shown in Fig.\ \ref{Fig: Laue}(b).

\clearpage

\section{Note 3: Supplemental information for methods}
Silver filled epoxy ohmic contacts were used for transport measurements.
The four probe DC current van der Pauw technique was employed to determine the resistivity of the samples.
The magnetic properties were studied using an MPMS-XL SQUID magnetometer from Quantum Design.
Temperature dependence of conductivity and magnetotransport measurements were made using a superconducting magnet based insert fitted into a liquid helium dewar. 
Transmission spectroscopy measurements were conducted with a microscope-type Fourier-transform infrared spectroscopy system (FT/IR-6600 and IRT-5200, JASCO) and a liquid-He cryostat (MicrsotatHe, Oxford Instruments).

In the band structure calculations, we used the norm-conserving pseudopotentials in the Pseudo-dojo project [31].
The equal $4\times 4\times 4$ $k$-point mesh was used for the charge density calculation, where the exact exchange energy was calculated with a supplementary $2\times 2\times 2$ $q$-point mesh.
The plane-wave energy cutoffs for the wave function and charge density were set to 70 Ry and 280 Ry, respectively.
The Wannier Hamiltonian was calculated with initial guesses with the Hg-$s$, $p$, Cr-$s$, $d$, and Se-$s$ and $p$ projectors.
We checked the agreement of the Wannier-interpolated energy levels and those in the self-consistent Kohn-Sham calculation at the equal mesh points.

\clearpage
\section{Note 4: Magnetotransport and magnetic properties}
\begin{figure}[hbt]
\includegraphics{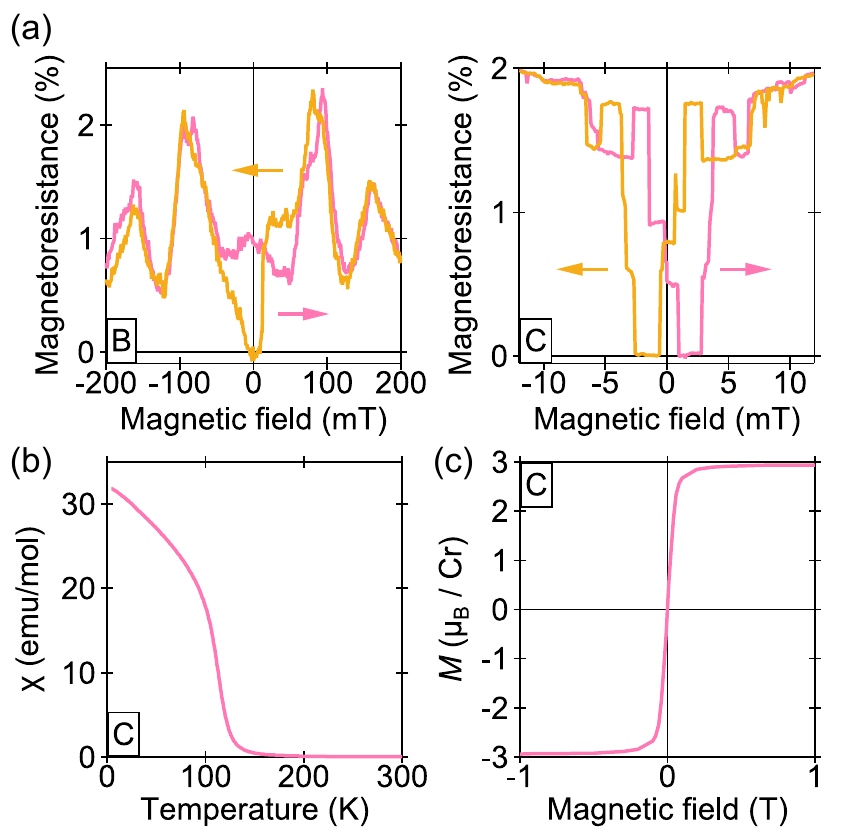}
\caption{\label{Fig: Magnetism} Magnetotransport and magnetic properties of $\mathrm{Hg}\mathrm{Cr}_2\mathrm{Se}_4$. (a) Magnetoresistance of crystals $B$ and $C$. The orange and pink arrows show the sweep direction of the magnetic field. (b) Temperature dependence of the magnetic susceptibility. (c) Magnetic field dependence of the magnetism.}
\end{figure}

During the magnetoresistance measurements, the temperature was fixed at 4.2 K. The magnetoresistance of $\mathrm{Hg}\mathrm{Cr}_2\mathrm{Se}_4$ exhibited anomalous behavior [Fig.\ \ref{Fig: Magnetism}(a)], likely due to a giant or tunneling magnetoresistance effect at the oppositely magnetized domain boundaries.

The magnetic susceptibility and magnetic moment were measured along the longer edge of a $\mathrm{Hg}\mathrm{Cr}_2\mathrm{Se}_4$ crystal.
During the magnetic susceptibility measurements, the magnetic field was fixed at 0.1 T, and during the magetic moment measurements, the temperature was fixed at 4.2 K.
Figure \ref{Fig: Magnetism}(b) shows that the ferromagnetic transition occurs at $T_C \simeq 110\ \mathrm{K}$.
The ferromangeitc moment is $2.95\ \mathrm{\mu}_\mathrm{B}/\mathrm{Cr}$ [Fig.\ \ref{Fig: Magnetism}(c)], which is consistent with the $S=3/2$ state of the $\mathrm{Cr}^{3+}$ ion in the crystal and the result of previous studies [13, 25].

\clearpage
\section{Note 5: Transmittance measurements}
\begin{figure}[hbt]
\includegraphics[width=160mm]{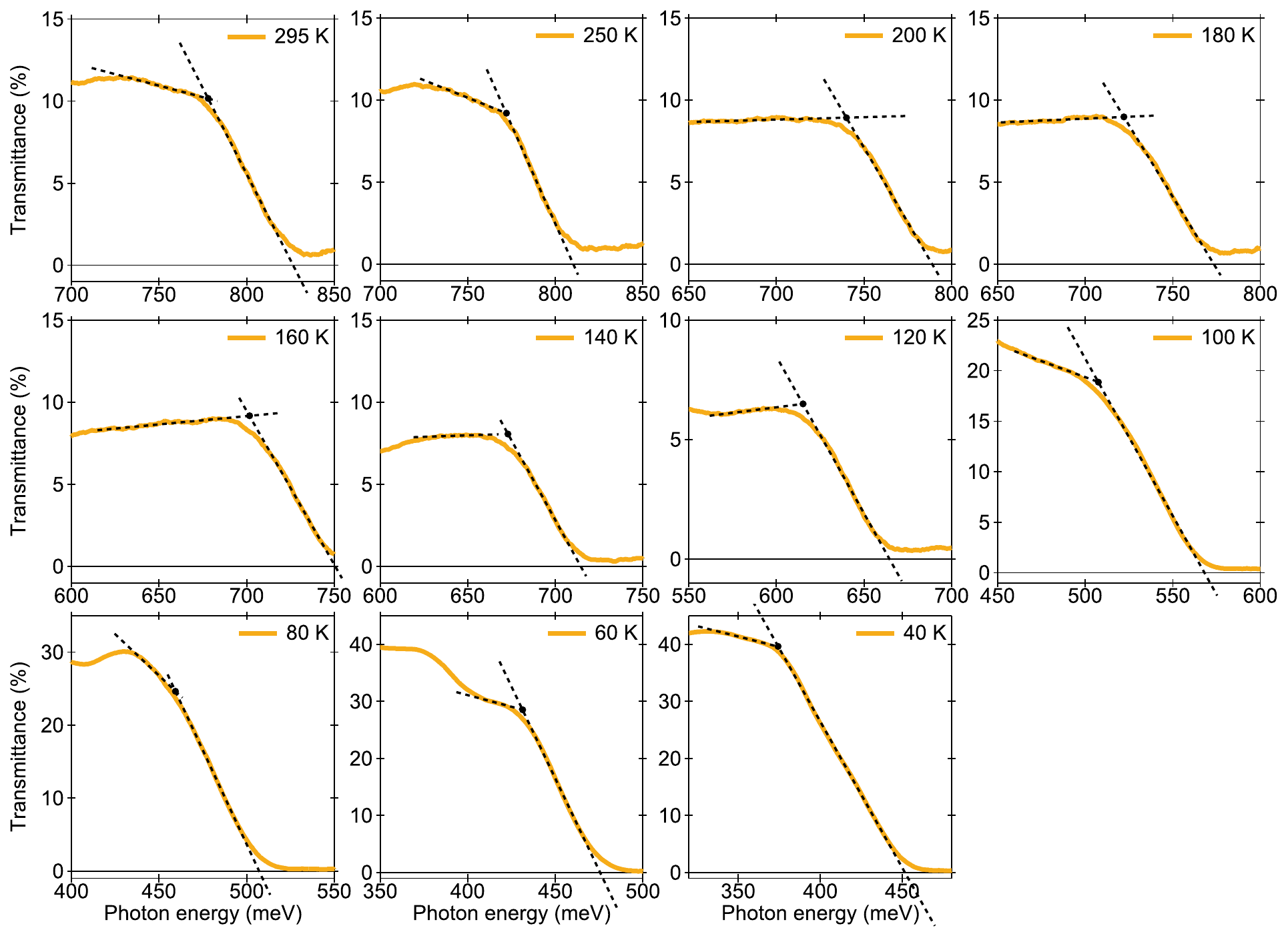}
\caption{\label{Fig: FTIR} Transmittance spectra. The two dashed lines in each panel show the fitting results to estimate the band gap.}
\end{figure}

Due to the limited number of single crystals, we performed transmittance measurements using a small piece of $\mathrm{Hg}\mathrm{Cr}_2\mathrm{Se}_4$.
Figure \ref{Fig: FTIR} shows the transmittance spectra near the band gap, measured while varying the cryostat temperature from 295 K to 40 K.
Generally, the band gap is obtained from the absorption edge calculated from the transmittance and reflectance spectra, as performed in Ref.\ [19].
However, the sample surface was not flat, and the thickness was not uniform, so we could not properly perform such an analysis.
Therefore, we estimated the band gap [Fig.\ 1(d)] as the point where the transmittance starts to drop.
At this point, the absorption coefficient is expected to increase rapidly.

\clearpage

\section{Note 6: Comparison between SX- and VUV-ARPES measurements}
\begin{figure}[hbt]
\includegraphics[width=160mm]{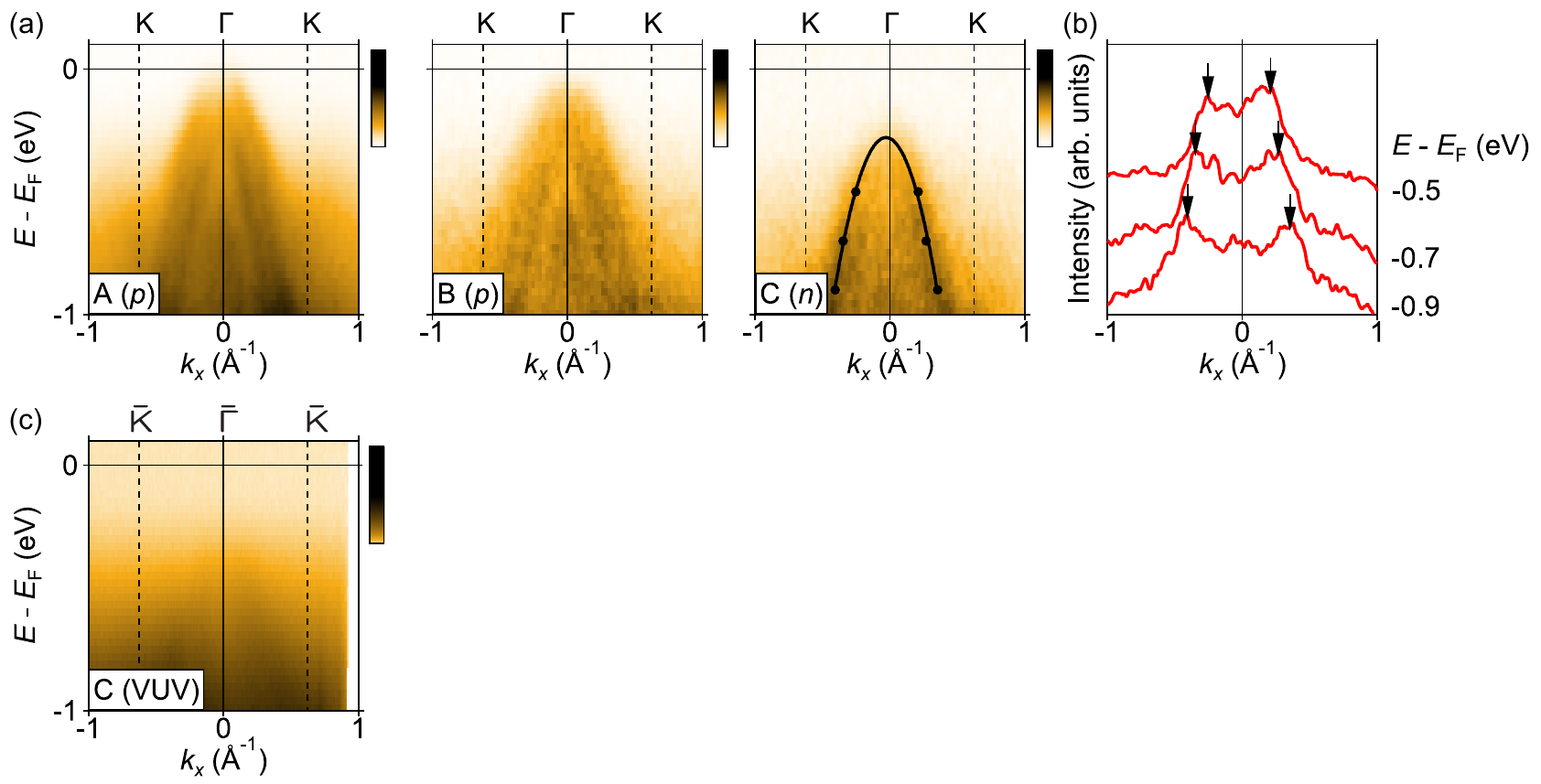}
\caption{\label{Fig: Ek} Band dispersions obtained by SX-ARPES ($h\nu=475\ \mathrm{eV}$) and VUV-ARPES ($60\ \mathrm{eV}$). (a) SX-ARPES measurement results for the three samples. The color maps in the first and third panels are the same as Figs.\ 2(c) and 2(d). The black dots represent the peak positions extracted from the momentum distribution curves (MDCs) in (b) and the black curve represents the fitting result of these points with a parabola. (b) MDCs of (a) at $-0.5\ \mathrm{eV}$, $-0.7\ \mathrm{eV}$, and $-0.9\ \mathrm{eV}$ from the Fermi level. The black arrows represent peak positions. (c) VUV-ARPES measurement result.}
\end{figure}

Figure \ref{Fig: Ek} compares band dispersions taken by SX- and VUV-ARPES for different samples.
The SX-ARPES measurements of crystal $B$ ($p$-type) exhibited hole bands at almost the same position as crystal $A$ ($p$-type).
We extracted the valence band top of $n$-type $\mathrm{Hg}\mathrm{Cr}_2\mathrm{Se}_4$ by fitting the hole band with a parabola [rightmost panel in Fig.\ \ref{Fig: Ek}(a)]. The obtained band top was 283 meV below the Fermi level.

We think the energy gap between the Fermi level and the valence band top in $n$-type $\mathrm{Hg}\mathrm{Cr}_2\mathrm{Se}_4$ does not contradict the optical band gap $\sim0.4\  \mathrm{eV}$ at 50 K.
That is because the valence or conduction band does not always cross the Fermi level in degenerate semiconductors.
In the ARPES study of Na-doped SnSe ($p$-type degenerate semiconductor) [32], the valence band top is about 100 meV off the Fermi level, and the tail of the peak crosses the Fermi level.
The resistivity of crystal $C$ below $T_C$ is still high, larger than $100\ \Omega\ \mathrm{cm}$, indicating that the density of states around the Fermi level is not so large.
Therefore, the conduction band bottom in $n$-type $\mathrm{Hg}\mathrm{Cr}_2\mathrm{Se}_4$ can be tens of milli-electron volts higher than the Fermi level. Combining these results, the energy gap between the valence band top and conduction band bottom can be around 300 - 400 meV, consistent with the optical band gap.

\begin{figure}[hbt]
\includegraphics{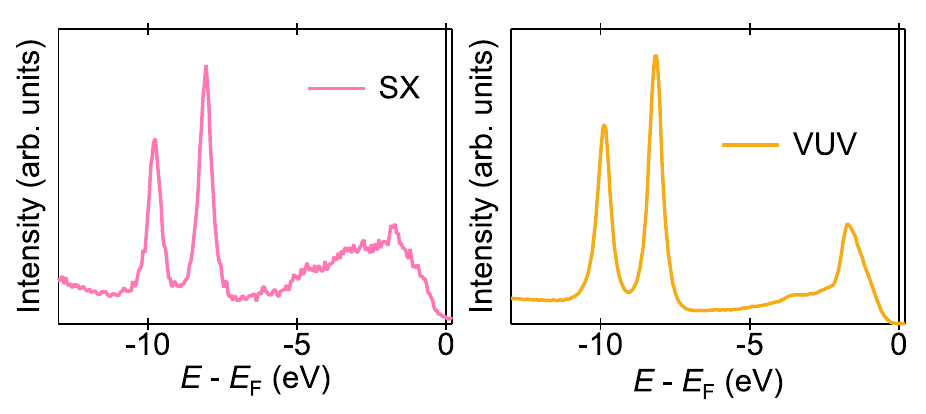}
\caption{\label{Fig: XPS} Hg $5d$ spectra observed by SX-ARPES ($h\nu=770\ \mathrm{eV}$) and VUV-ARPES ($60\ \mathrm{eV}$).}
\end{figure}

While SX-ARPES measurements [Fig.\ \ref{Fig: Ek}(a)] detected hole bands, we could not see a band in VUV-ARPES measurements of crystal $C$ [Fig.\ \ref{Fig: Ek}(c)].
This difference is not due to the cleave failure because we observed similar Hg $5d$ core level spectra in both SX- and VUV-ARPES [Fig.\ \ref{Fig: XPS}].

\clearpage
\section{Note 7: DFT band dispersions with different $\alpha$ values}

\begin{figure}[hbt]
\includegraphics{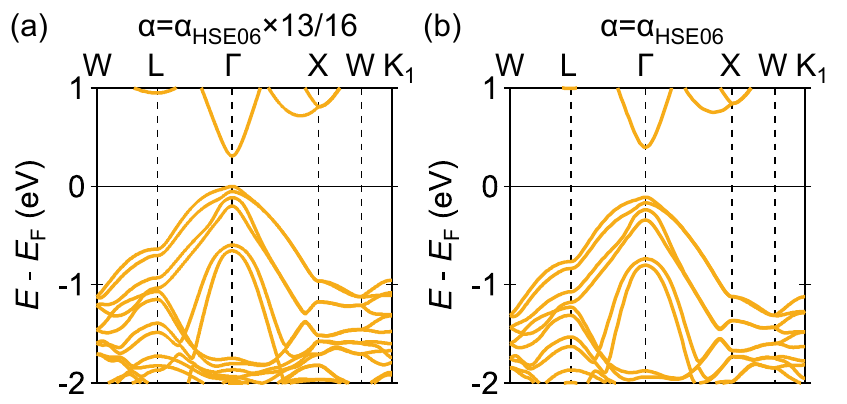}
\caption{\label{Fig: Overall_alpha} Calculated band dispersions with different $\alpha$ values. (a) Band dispersion with $\alpha=\alpha_\mathrm{HSE06}\times 13/16$, which is the same as in Fig.\ 3(b). (b) Band dispersion with $\alpha=\alpha_\mathrm{HSE06}$.}
\end{figure}

Figure \ref{Fig: Overall_alpha} compares band dispersions calculated with different $\alpha$ values.
Calculations with $\alpha=\alpha_\mathrm{HSE06}\times13/16$ exhibited a band gap of 0.31 eV [Fig.\ \ref{Fig: Overall_alpha}(a)], approximately the same as the experiment at low temperatures [19].
On the other hand, the band dispersion with the standard HSE06 value shows a larger band gap (0.51 eV) [Fig.\ \ref{Fig: Overall_alpha}(b)].
However, the overall band structure is kept unchanged with changing $\alpha$.

\section{Note 8: Symmetry lowering due to the ferromagnetic moments}
\begin{figure}[hbt]
\includegraphics{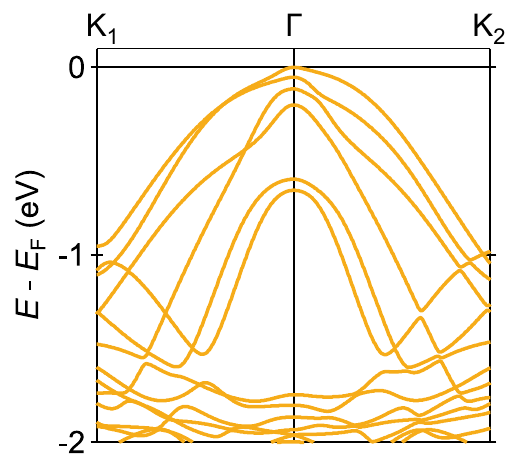}
\caption{\label{Fig: KxyGKyz}Band dispersion along the $K_1\Gamma$ and $\Gamma K_2$ paths. The $K_1$ and $K_2$ points are defined in Fig.\ 1(b).}
\end{figure}

The ground state obtained by DFT calculations exhibits ferromagnetism along the vertical direction in Fig.\ 1(b).
The symmetry lowering due to the ferromagnetic moment changes the $K_1$ and $K_2$ points inequivalent.
However, the effect of the symmetry lowering on the band dispersion is small [Fig.\ \ref{Fig: KxyGKyz}].

\section{Note 9: Resonant PES measurements}
\begin{figure}[hbt]
\includegraphics{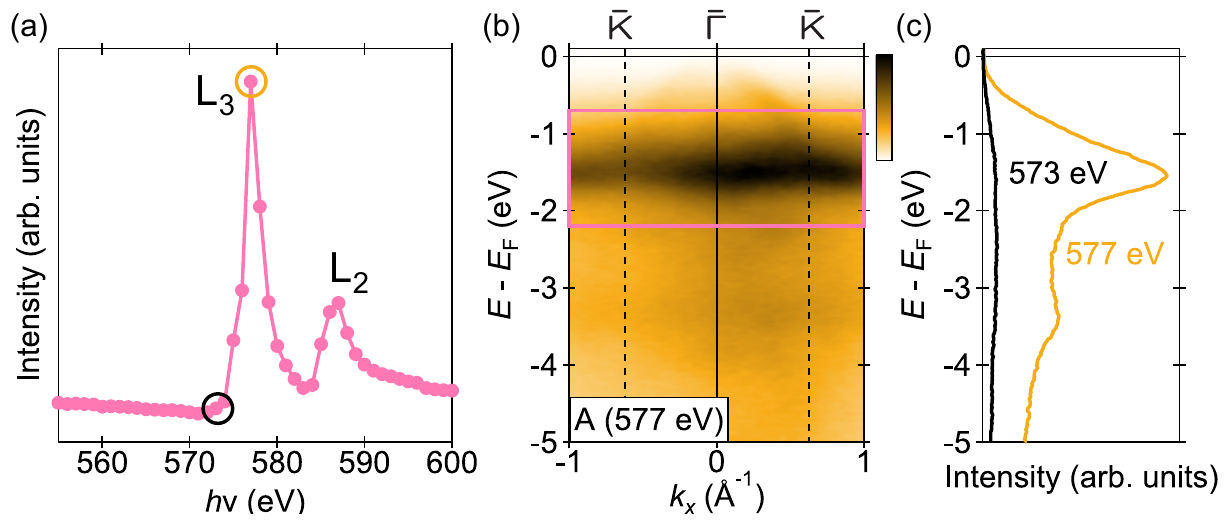}
\caption{\label{Fig: Resonant}Resonant PES measurements. (a) $h\nu$ dependence of the photoemission intensity near the $\mathrm{L}_2$ and $\mathrm{L}_3$ edges of Cr. The orange and black circles represent the energies on- and off-resonance, respectively. (b) Band dispersion at the energy on Cr resonance (577 eV). The pink rectangle represents the area to calculate the integrated intensity in (a). (c) Integrated energy distribution curves at on (577 eV) and off (573 eV) resonance.}
\end{figure}

Figure \ref{Fig: Resonant}(a) shows two peak structures of the integrated photoemission intensity near the $\mathrm{L}_3$ (577 eV) and $\mathrm{L}_2$ (587 eV) edges.
The energy distribution curve taken at the energy on resonance has enhanced intensity between $-4\ \mathrm{eV}$ and $-1\ \mathrm{eV}$ from the Fermi level [Fig.\ \ref{Fig: Resonant}(c)].
This area corresponds to the density distribution of the Cr $3d$ orbital in DFT calculations [Fig.\ 3(d)].

\section{Note 10: Photoemission from the rough surface region and deeper region}
\begin{figure}[hbt]
\includegraphics{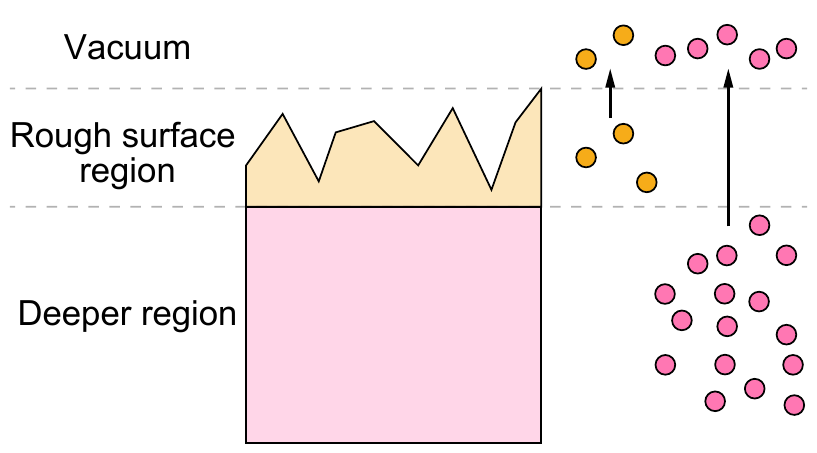}
\caption{\label{Fig: Photoemission_sch} Schematic of the photoemission from the rough surface region and deeper region.}
\end{figure}

Figure \ref{Fig: Photoemission_sch} explains our interpretation of the observed spectra based on the three-step model.
Since the penetration depth of soft-x-ray is much longer than a few nanometers, photoelectrons are generated below the rough surface region.
However, the number of generated photoelectrons in the rough surface region is expected to be much smaller than that in the deeper region due to the broken in-plane translation symmetry in the rough surface region.
Since the photoemission probability is roughly determined by the matrix element of the plane wave and the ground state wave function, the incoherent electronic structure in the rough surface region makes the photoemission probability very small and broad in the momentum space.
Therefore, although many photoelectrons generated in the deeper region are scattered, we observed the unscattered photoelectrons forming the hole bands related to the bulk electronic structure.
The photoelectrons from the surface region and the scattered electrons form the background of the ARPES spectra.

\section{Note 11: Refraction in the photoemission process and the simulation for randomly oriented planes}

\begin{figure}[hbt]
\includegraphics{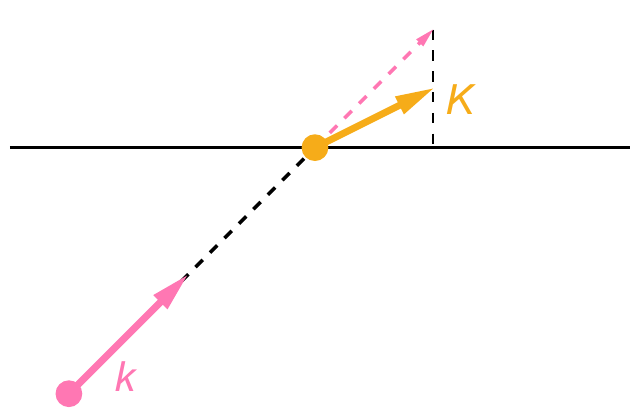}
\caption{\label{Fig: Refraction_sch} Schematic of the refraction in the escape process of the three-step photoemission model.}
\end{figure}

Figure \ref{Fig: Refraction_sch} represent the concept of the refraction in the last step of the three-step model [1].
The kinetic energy of a photoelectron $E_\mathrm{kin}$ satisfies
\begin{equation}
E_\mathrm{kin}=\frac{\hbar^2 |\mathbf{K}|^2}{2m}=\frac{\hbar^2 |\mathbf{k}|^2}{2m}-V_0,
\end{equation}
where $V_0$ is the inner potential of the material.
Therefore $\mathbf{K}$ and $\mathbf{k}$ have different length.
Since the in-plane components of the wave vector should be conserved, the out-of-plane component is changed.

In the simulations in both SX and VUV regions, the band dispersion is free-electron-like around the $\Gamma$ point in the periodic zone scheme.
Randomly oriented flat surfaces were obtained from randomly oriented normal vectors $\mathbf{v}=(x,\ y,\ z)$ satisfying $|\mathbf{v}|=1$ and $z>0$.
The $k_z$ coordinates were set to $11\times2\sqrt{3}\pi/a$ for SX and $4\times2\sqrt{3}\pi/a$ for VUV, respectively; $a=10.7\ \text{\AA}$ is the lattice constant of the cubic unit cell.

\begin{figure}[hbt]
\includegraphics{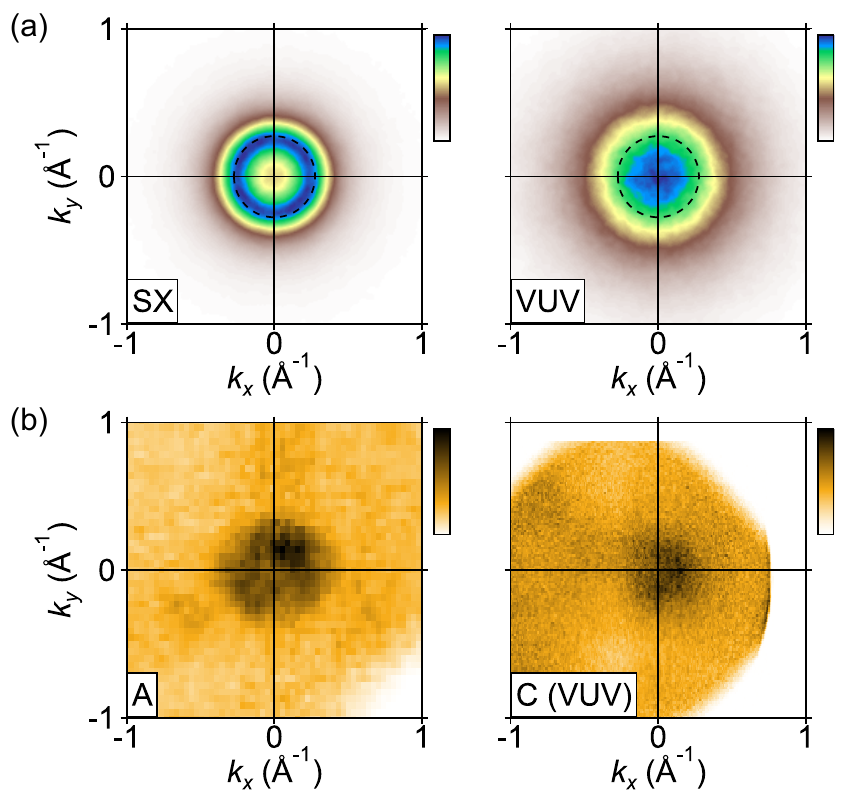}
\caption{\label{Fig: Refraction_FS} Simulated and experimental constant energy maps at 0.4 eV below the Fermi level for SX and VUV-ARPES.}
\end{figure}

Figure \ref{Fig: Refraction_FS} compares simulated and experimental constant energy maps.
In the SX region, the simulated dispersion becomes circle-like, consistent with the original dispersion [dashed circle in the top panels].
SX-ARPES measurements also show a less clear circle-like dispersion.
On the other hand, the simulated spectra in the VUV region do not have a circle dispersion.
The VUV-ARPES measurements showed similar behavior.

\begin{figure}[hbt]
\includegraphics{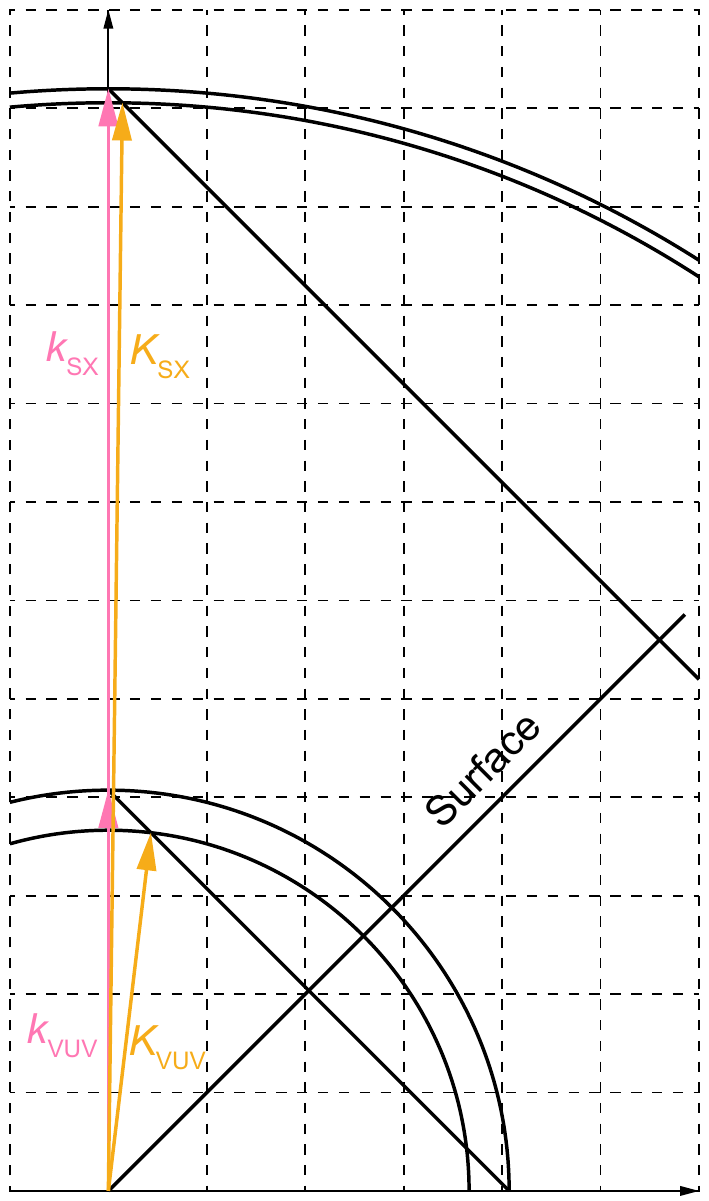}
\caption{\label{Fig: Refraction_45deg} Wave vectors before and after the refraction by the $45^{\circ}$-tilted surface. Original wave vectors $k_\mathrm{SX}$ and $k_\mathrm{VUV}$ correspond to the $\Gamma$ points. The grid is $1\ \text{\AA}^{-1}$.}
\end{figure}

Figure \ref{Fig: Refraction_45deg} describes the difference between the refraction in SX and VUV regions.
Although $k_\mathrm{SX}$ and $k_\mathrm{VUV}$ represent the same $\Gamma$ points, the refracted wave vector $K_\mathrm{VUV}$ is more tilted than $K_\mathrm{SX}$.